# Magnetic behavior of nano crystals of a spin-chain system, $Ca_3Co_2O_6$: Absence of multiple steps in the low-temperature isothermal magnetization

Niharika Mohapatra, Kartik K Iyer, Sitikantha D Das, B.A. Chalke, S.C. Purandare, and E.V. Sampathkumaran[*]

*Tata Institute of Fundamental Research, Homi Bhabha Road, Colaba, Mumbai – 400005, India.*

**Abstract**

We report that the major features in the temperature dependence of *dc* and *ac* magnetization of a well-known spin-chain compound, $Ca_3Co_2O_6$, which has been known to exhibit two complex magnetic transitions due to geometrical frustration (one near 24 K and the other near 10 K), are found to be qualitatively unaffected in its nano materials synthesized by high-energy ball-milling. However, the multiple steps in isothermal magnetization - a topic of current interest in low-dimensional systems - known for the bulk form well below 10 K is absent in the nano particles. We believe that this finding will be useful to the understanding of the 'step' magnetization behavior of such spin-chain systems.

PACS numbers: 75.60.Ej; 75.40.Cx; 75.50.-y; 75.50.Lk

The existence of the plateaus in the isothermal magnetization (*M*) of spin-chain systems has been attracting considerable attention of condensed matter theorists as well as experimentalists [1]. In this respect, the spin-chain compound, $Ca_3Co_2O_6$, crystallizing in the $K_4CdCl_6$-type rhombohedral structure is of great interest in the current literature, as there is one step in *M* in an intermediate temperature (*T*) range (~ 10 - 24 K) at one third of saturation magnetization ($M_s$), but there are multiple steps at much lower temperatures at an equal spacing of 12 kOe [2-4]. It is now well-known that the ferromagnetic spin-chains are arranged in a triangular fashion in the basal plane with an inter-chain antiferromagnetic interaction leading to geometrical frustration in this compound; as a result, 2 out of 3 chains order antiferromagnetically and the third one remains incoherent (in the range 10 – 24 K) or undergoes a complex ordering (below 10 K). With this scenario, the plateau at $M_s$/3 for the range 10-24 K is understandable in terms of a field-induced ferrimagnetic alignment [2], but it is difficult to explain the step-behavior at lower temperatures. Various explanations have been proposed [4-10] in the literature for the origin of multi-step behavior below 10 K, for example, (i) magnetic field (*H*) induced transitions between different spin states [4,7], and (ii) quantum tunneling effects similar to that known for molecular magnets and a complex relaxation process [5, 6].

Apart from above magnetization behavior, this compound provides a unique opportunity to understand geometrically frustrated magnetism of spin-chains. The magnetic state in the range 10 – 24 K is called "partially disordered antiferromagnetism (PDA)". Interesting anomalies have been found in many spectroscopic and bulk studies, including magnetodielectric coupling and thermopower from the applications point of view [11-21]. While all these studies have been carried out on bulk single crystals or polycrystals, there is not much work on the nano crystalline form. There were two reports [22] on the isothermal *M* behavior at 10 K on the films grown on different substrates by pulsed laser ablation method and the results appear to differ. In addition, it is not clear how these 'steps' in *M(H)* behave at further low temperatures in these films. In order to advance the understanding of the properties of this compound in general, it is important to clarify whether the magnetic properties are preserved down to nanoscale, particularly noting that the magnetic correlation length scale along *c*-axis has been experimentally determined to be of the order 550 nm along c-axis, whereas it is of the order 18 nm along basal plane [14, 15]. We have therefore synthesized the fine particles in nanometer range by high-energy ball-milling method [24] and studied its magnetization behavior, the results of which are reported here.

The polycrystalline samples in the bulk form (called '***B***') have been prepared by a conventional solid state reaction route [11] and characterized by x-ray diffraction (XRD) (Cu $K_\alpha$) to be single phase (within the detection limit of 2%). The compound, after characterizing by magnetic studies as well, were milled for 8 hours in a planetary ball-mill (Fritsch pulverisette-7 premium line) operating at a speed of 500 rpm in a medium of toluene to attain nanosized (called '***N***') particles. Tungsten carbide vials and balls of 5 mm diameter were used with a balls-to-material mass ratio of 10:1. To characterize the specimens, apart from x-ray diffraction, scanning electron microscope (SEM) and transmission electron microscope (TEM, Tecnai 200 kV) were employed. The *dc M(T)* measurements (in the range 1.8-300 K) were performed employing a commercial superconducting quantum interference device (SQUID, Quantum Design) and the same magnetometer was employed to take *ac* susceptibility ($\chi_{ac}$) data. *M(H)* (up to 120 kOe) measurements at selected temperatures were carried out employing a commercial vibrating sample magnetometer (VSM, Oxford Instruments).

Figure 1a shows the XRD pattern for the milled specimen. All reflections expected are present in the milled specimen and there is no change in the shape of the background with respect to that of the bulk. There is negligible change in the lattice constants [For **B**: $a$= 9.076(2) Å, $c$= 10.379(2) Å; for **N**, $a$= 9.069(2)Å, $c$= 10.393(2) Å)] obtained by Reitveld fitting. The pattern for the milled sample is more broadened, attributable to a decrease in particle size, as shown in the inset of figure 1a. An idea of the average particle size (about 50 nm) could be inferred from the width of the most intense line (after subtracting instrumental line-broadening) employing Debye-Scherrer formula. Attempts to eliminate corrections for strain effects employing Williamson-Hall plots were not successful, as such plots are not found to be linear, possibly due to a large spread in the particle sizes. In order to get a better idea about the particle size, we have employed SEM and TEM. According to the SEM images, shown in figure 1b, the particles attained nanometer dimensions (typically below 100 nm) with significant agglomeration. We have isolated some of the particles by ultrasonification in alcohol and obtained the bright-field TEM images. These images, shown in figure 1c, reveal that these particles are rod-shaped with a typical length of a few hundred nanometer and a width of less than 50 nm. It is interesting to note that the ball-milling conditions employed yields nano rods, which otherwise could be obtained by laser ablation only on Si(100) substrate. High-resolution TEM images (see figure 1d, for instance for (300) plane) reveal well-defined lattice planes, thereby confirming that the nanospecimens are crystalline, and not amorphous. We have also obtained the selected area electron diffraction pattern (see figure 1e) and all the diffraction rings are indexable to the compound under investigation, thereby confirming that the nano specimen correspond to the parent compound and these are polycrystalline; the appearance of some bright spots along the diffraction rings reveal that the particles are highly textured. Polycrystalline nature of the rods were also confirmed by dark field images (not shown here).

The results of *dc* and *ac* magnetization as a function of temperature measured in the presence of various magnetic fields ($H$= 100 Oe, 5 kOe, 10 kOe and 50 kOe) are shown in figures 2 and 3. For comparison, a curve obtained in a field of 5 kOe for the zero-field-cooled condition (ZFC, from 100 K) for the bulk form [3, 4, 11] are also shown. From the mainframe of figure 2, it is obvious that there is a sudden upturn in magnetization measured in low fields (<<50 kOe) near ($T_1$=) 24 K, with a peak in the ZFC curve at a lower temperature ($T_2$ ~ 8 to10 K). Application of high fields (say, 50 kOe) broadens the features at these transitions. These features, similar to those observed for the bulk form, imply that $T_1$ and $T_2$ are essentially unaffected by reducing the particle size. As a characteristic feature of PDA ordering between $T_1$ and $T_2$, the *dc* χ curves are strongly field-dependent similar to that known for **B**. In order to address whether there is any change in the magnetic moment on Co, we have performed *dc* χ measurements up to 300 K in the presence of a field of 50 kOe. As shown in the inset of figure 2, the plot of inverse χ versus *T* is linear over a wide T-range (100 - 300 K), and the effective moment obtained from the linear region is about 5.6 ± 0.05 $\mu_B$ per formula unit, which is essentially the same as that of bulk. This implies that the electronic configuration of Co remains unaffected as one goes from bulk to nano form. However, the sign of paramagnetic Curie temperature (~ -20 K) for **N** is negative, whereas it is positive for **B** ( ~ 35 K), as though the net dominating correlation is antiferromagnetic for **N**, precise reason of which is not clear to us at the moment. Otherwise, the above results reveal that the gross magnetic features of **B** are retained in **N**.

The $\chi_{ac}$ curves for **N** also look similar (Fig. 3) to that of **B** in the sense that there is a peak in both real (χ') and imaginary (χ") parts, exhibiting a huge frequency (ν) dependence, with the χ'-peak temperature moving from 10 K for 1 Hz to 16 K for 1.339 kHz, implying that the spin-dynamics are undisturbed when the particle size is reduced in this temperature range. We had earlier reported [20] that Arrhenius relation is obeyed; we estimate the activation energy from this relationship to be about 145 K, which is in good agreement with Hardy et al [6]. An upward curvature in the real part of $\chi_{ac}$, reported by Hardy et al [Ref. 6] in single crystals at very low temperatures (<5 K) and attributed to an intriguing change in spin dynamics, could be observed in the specimen *N* as well.

We now look at the *M(H)* behavior (see figure 4), which were taken for the ZFC condition of the specimens. We have chosen one temperature well below $T_2$ and another between $T_1$ and $T_2$, *viz.,* 1.8 and 15 K, to drive the messages. We note that, for *N*, the magnetic moment does not get saturated even at fields as high as 120 kOe and the value of the magnetic moment at 120 kOe is significantly lower (nearly half) than that of the specimen *B*. Also, the *M(H)* curve (at 1.8 K) is hysteretic. These features may support the inference from the paramagnetic Curie temperature that antiferromagnetic correlation tend to dominate when the particle dimension is reduced. With respect to steps in *M(H)*, as mentioned in the introduction, for the specimen *B*, for *T*=15 K, there is a plateau at nearly $M_s/3$ and this feature appears manifesting itself as a distinct upturn in *M* near 35 kOe for *N*, pointing to the fact that PDA order is maintained at this temperature in the nanoform as well. For *T*= 1.8 K, the multiple steps, clearly appearing in single crystals [4], are usually weakened in polycrystalline bulk samples and appear as a change of slope of *M(H)* plot. The derivative of *M* with respect *H* reveals minima at these steps in *M(H)*, as shown in the inset of figure 4. These steps are irreversible at this temperature. It is clear from figure 4 that the features are absent for the specimen *N* and *dM/dH* undergoes negligible variation (till the field of interest of 60 kOe) *without exhibiting any minimum.* This is interesting, particularly noting that the behavior above $T_2$ for *N* is essentially unaffected with respect to that for *B*.

As mentioned in the introduction, different concepts have been proposed for the multi-step *M(H)* behavior and characteristic *ac* and *dc* magnetization behavior and magnetic relaxation curves have been established for the single crystals in the past [4,5]. In order to explore whether these characteristic features are still retained for *N*, we took additional magnetic data for *N*. We have obtained the *M(H)* curves with VSM for different rates of change of *H* (100 Oe/min to 10 kOe/min) and it was found that the behavior is independent of the rate of change of field. In fact, the curves for varying sweep rates of field are found to lie one over the other. This situation is different from that reported for the bulk single crystals [5], in which case, the steps get more prominent with an increasing rate of change of *H*. In order to understand the behavior of spin-relaxation time ($\tau$), we have measured with SQUID the imaginary part ($\chi''$) of $\chi_{ac}$ as a function of $\nu$ (0.03 Hz – 1 kHz) at several temperatures below 14 K after cooling the sample in zero field to the desired temperature and obtained $\tau$ (= $1/2\pi\nu_m$) from the knowledge of $\nu_m$ at which $\chi''$ exhibits a peak. It is straightforward to conclude from the comparison of the curves in figure 5a with those reported for the single crystals by Hardy et al in figure 2b in Ref. 6 that the $\tau$ values fall to comparatively much higher values with decreasing temperature; for instance, a distinct peak in $\chi''$ could be seen near 0.3 Hz for *T*= 2.25 K in single crystals, whereas for *N*, no peak could be observed above 0.03 Hz even at 4 K. Since we can not measure $\chi''$ in the $\nu$-range below 0.03 Hz with our magnetometer, we have inferred the trend in $\tau$ for $T \leq 6$ K by isothermal remnant magnetization ($M_{IRM}$) behavior (measured with VSM). For this purpose, following the procedure adopted for single crystals by Maignan et al (Ref. 6), after the application of a magnetic field of a high field (say, 70 kOe), $M_{IRM}$ in zero field was tracked as a function of time (*t*). $M_{IRM}$ decays with *t* (see figure 5b) and, by fitting the curve to a stretched exponential of the form, $M_{IRM} = a + b \exp[-(t/\tau)^{0.5}]$ (where *a* and *b* are constants), we have estimated the values of $\tau$.

Qualitatively speaking, a combined look at the values of $\tau$ determined from both the methods (see figure 5a, inset) reveals that $\tau$ exhibits thermally activated behavior well above 6 K with values less than few seconds and there is a crossover regime near $T_2$ at which $\tau$ increases to several minutes increasing monotonically with T for instance, from ~ 60 sec at 6 K to ~ 1400 sec at 2 K. From these observations, it is concluded that the constancy of the value seen for the single crystal below 6 K is however absent for *N*. These findings suggest that the concepts of quantum tunneling and multiple-spin states proposed for bulk form need not be considered for *N*. This observation raises a question whether the multiple-step *M(H)* feature for the bulk form is characterized by a length scale much larger than the dimensions of the nano particle, particularly noting that the magnetic correlations lengths (see the introduction) are of comparable magnitudes as those of *N*. Alternatively, a possible difference in the magnetic structure (below 10 K) between *B* and *N* could also be responsible for the modification of the low-temperature magnetization behavior.

We have made another interesting finding in the *dc* $\chi$ plot. It is known [3, 4] that ZFC and FC curves tend to bifurcate near $T_2$ in the bulk form. A careful look at the data for **N** measured with low fields, say for *H*= 100 Oe, revealed (see figure 2, inset *b*) that the bifurcation sets in at a much higher temperature, well above $T_1$ (near 45 K), similar to the behavior noted for nano rods and thin films [22]. This finding seems to endorse our previous claim [17, 21] with respect to additional interesting physics in the higher temperature range.

Summarizing, the central point of emphasis is that the nano particles of the geometrically frustrated spin-chain compound, $Ca_3Co_2O_6$, synthesized by high-energy ball-milling, do not show multiple step isothermal magnetization behavior, noted for the bulk form of this compound at low temperatures. However, other qualitative features in the magnetic susceptibility remain essentially unaltered compared to the bulk. We believe that this work would motivate an extension of such studies to the nano particles of other spin chain systems to clarify pertinent issues. Another finding we made from the low-field magnetic susceptibility data is that there exists another magnetic anomaly above 40 K. Finally, this work provides a route to make large quantities of nano particles of this spin-chain oxide and its derivatives in stable form to enable further studies and potential applications.

We thank N.R. Selvi Jawaharlal Nehru Center for Advanced Scientific Research, Bangalore, India, for SEM data.

*E-mail address: sampath@mailhost.tifr.res.in

1. See, for example, T. Vekua, D.C. Kabra, A. Dobry, C. Gazza, and D. Poilblanc, Phys. Rev. Lett. 96, 117205 (2006).
2. S. Aasland, H. Fjellvag, and B. Hauback, Solid State Commun. 101, 187 (1997).
3. H. Kageyama, K. Yoshimura, K. Kosuge, H. Mitamura, and T. Goto, J. Phys. Soc. Jpn. 66, 1607 (1997).
4. A. Maignan, C. Michel, A.C. Masset, C. Martin, and B. Raveau, Eur. Phys. J. B 15, 657 (2000).
5. V. Hardy, M.R. Lees, O.A. Petrenko, D. McK Paul, D. Flahaut, S. Hebert, and A. Maignan, Phys. Rev. B 70, 064424 (2004).
6. V. Hardy, D. Flahaut, M.R. Lees, and O.A. Petrenko, Phys. Rev. B 70, 214439 (2004); A. Maignan, V. Hardy, S. Hebert, M. Drillon, M.R. Lees, O. Petrenko, D. McK Paul, and D.Khomskii, J. Materials Chem. 14, 1231 (2004).
7. H. Wu, M.W. Haverkort, Z. Hu, D.I.Khomskii, and L.H. Tjeng, Phys. Rev. Lett. 95, 186401 (2005)
8. Yu. B. Kudasov, A.S. Korshunov, V.N. Pavlov, and D.A. Maslov, Phys. Rev. B 78, 132407 (2008).
9. X. Yao, S. Dong, H. Yu, and J.Liu, Phys. Rev. B 74, 134421 (2006).
10. R. Sato, G. Martinez, M.N. Baibich, J.M. Florez, and P. Vargas, arXiv:0811.4772.
11. S. Rayaprol, K. Sengupta. And E.V. Sampathkumaran, Solid State Commun. 128, 79 (2003); E.V. Sampathkumaran, N. Fujiwara, S. Rayaprol, P.K. Madhu, and Y. Uwatoko, Phys. Rev. B 70, 014437 (2004); E.V. Sampathkumaran, Z. Hiroi, S. Rayaprol, and Y. Uwatoko, J. Magn. Magn. Mater. 284, L7 (2004).
12. K. Takubo, T. Mizokawa, S. Hirata, J.-Y. Son, A. Fujimori, D. Topwal, D.D. Sarma, S. Rayaprol, and E.V. Sampathkumaran, Phys. Rev. B 71, 073406 (2005).
13. T. Burnus, Z. Hu, M.W. Haverkort, J.C. Cezar, D. Flahaut, V. Hardy, A. Maignan, N.B. Brookes, A. Tanaka, H.H. Hsich, H.-J. Lin, C.T. Chen, and L.H. Tjeng, Phys. Rev. B 74, 245111 (2006).
14. A. Bombardi, C. Mazzoli, S. Agrestini, and M.R. Lees, Phys. Rev. B 78, 100406(R) (2008); S. Agrestini, C. Mazzoli, A.Bombardi, and M.R. Lees, Phys. Rev. B 77, 140403(R) (2008).
15. S. Agrestini, L.C. Chapon, A. Dauod-Almadine, J. Schefer, A. Gukasov, C.Mazzoli, M.R. Lees, and O.A. Petrenko, Phys. Rev. Lett. 101, 097207 (2008).
16. S. Takeshita, J. Arai, T. Goko, K. Nishiyama, and K. Nagamine, J. Phys. Soc. Jpn. 75, 034712 (2006); J. Sugiyama, H. Nozaki, J.H. Brewer, E.J. Ansaldo, T. Takami, H. Ikuta, and U. Mizutani, Phys. Rev. B 72, 064418 (2005).
17. P.L. Paulose, N.Mohapatra, and E.V. Sampathkumaran, Phys. Rev. B 77, 172403 (2008).
18. N. Bellido, C. Simon, and A. Maignan, Phys. Rev. B 77, 054430 (2008).
19. M. Mikami, R. Funahashi, M. Yoshimura, Y. Mori, and T. Sasaki, J. Appl. Phys. 94, 6579 (2003).
20. S. Rayaprol, Kausik Sengupta and E.V. Sampathkumaran, Proc. Indian Acad. Sci. (Chem. Sci), 115, 553 (2003).
21. R. Bindu, K. Maiti, S. Khalid, and E.V. Sampathkumaran, Phys. Rev. B 79, 094103 (2009).

22. P.L. Li, X.Y. Yao, F. Gao, C. Zhao, K.B. Yin, Y.Y. Weng, J.-M. Liu, and Z.F. Ren, App. Phys. Lett. 91, 042505 (2007); R. Mouben, A. Bouaine, C. Ulhaq-Bouillet, G. Schmerber, G. Versini, S. Barre, J.L. Loison, M. Drillon, S. Colis, and A. Dinia, App. Phys. Lett. 91, 172517 (2007).
23. C. Suryanarayana, Prog. Mater. Sci. 46, 1 (2001).

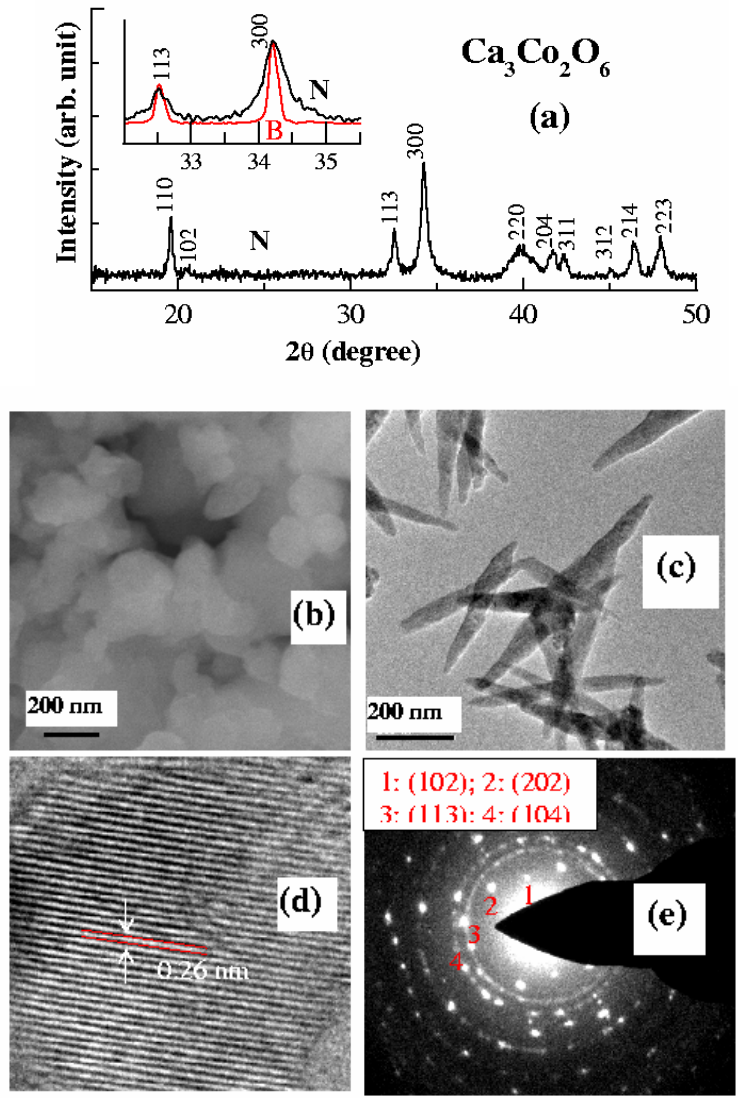

Figure 1:
(color online) **(a)** X-ray diffraction pattern (Cu K$_\alpha$), **(b)** SEM images, **(c)** TEM images of nano rods, (d) high-resolution TEM showing (300) lattice planes, and (e) selected area diffraction pattern obtained by TEM with indexing of four innermost diffraction rings, for the ball-milled specimens of $Ca_3Co_2O_6$. In **(a),** the x-ray diffraction pattern for two lines are compared for the bulk and nano particles to show line-broadening (after normalizing to respective peak heights).

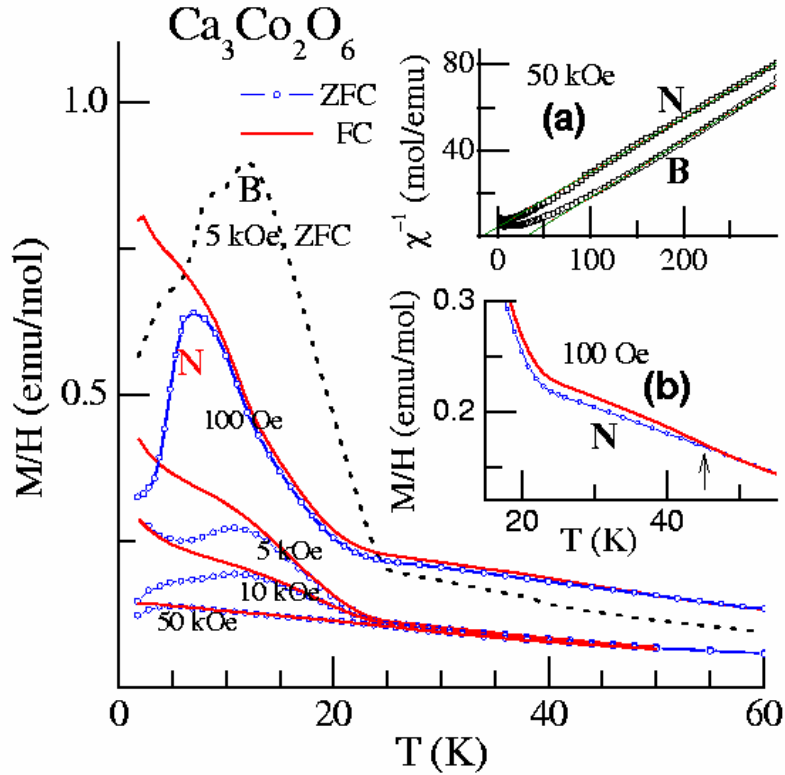

Figure 2:
(color online) Magnetization (*M*) divided by magnetic field (*H*), measured in the presence of several fields below 60K for the nano particles (*N*) of $Ca_3Co_2O_6$. The ZFC-curve obtained in 5 kOe for the bulk specimen (*B*) is also shown for comparison. In inset *(a)*, inverse susceptibility obtained in a field of 50 kOe is plotted for both the specimens and the straight lines in these cases are obtained by Curie-Weiss fitting of the data above 100 K. In inset *(b),* the magnetic susceptibility curves obtained in a field of 100 Oe for the zero-field-cooled and field-cooled conditions of the nano specimen are shown in an expanded form to highlight apparent bifurcation around 45 K. The lines through the data points serve as guides to the eyes.

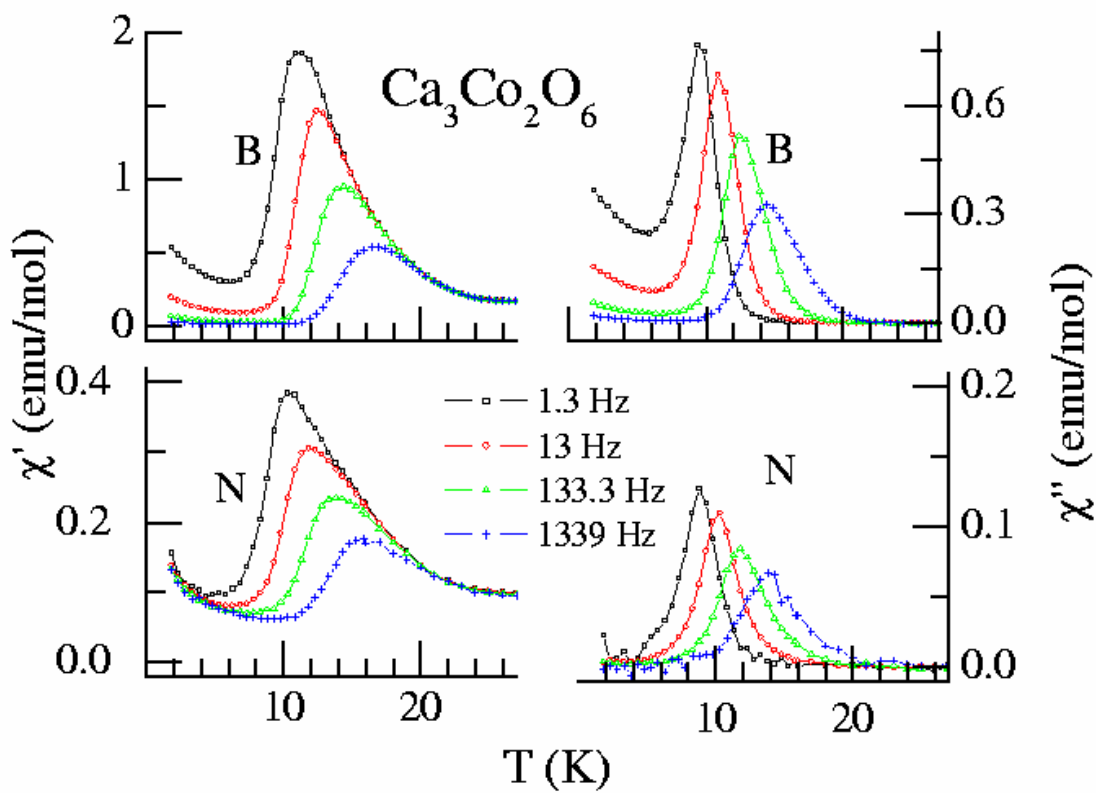

Figure 3:
(color online) Real ($\chi'$) and imaginary ($\chi''$) parts of *ac* susceptibility measured at various frequencies (with a *ac* field of 1 Oe) for bulk and nanoparticles of $Ca_3Co_2O_6$. The lines through the data points serve as guides to the eyes.

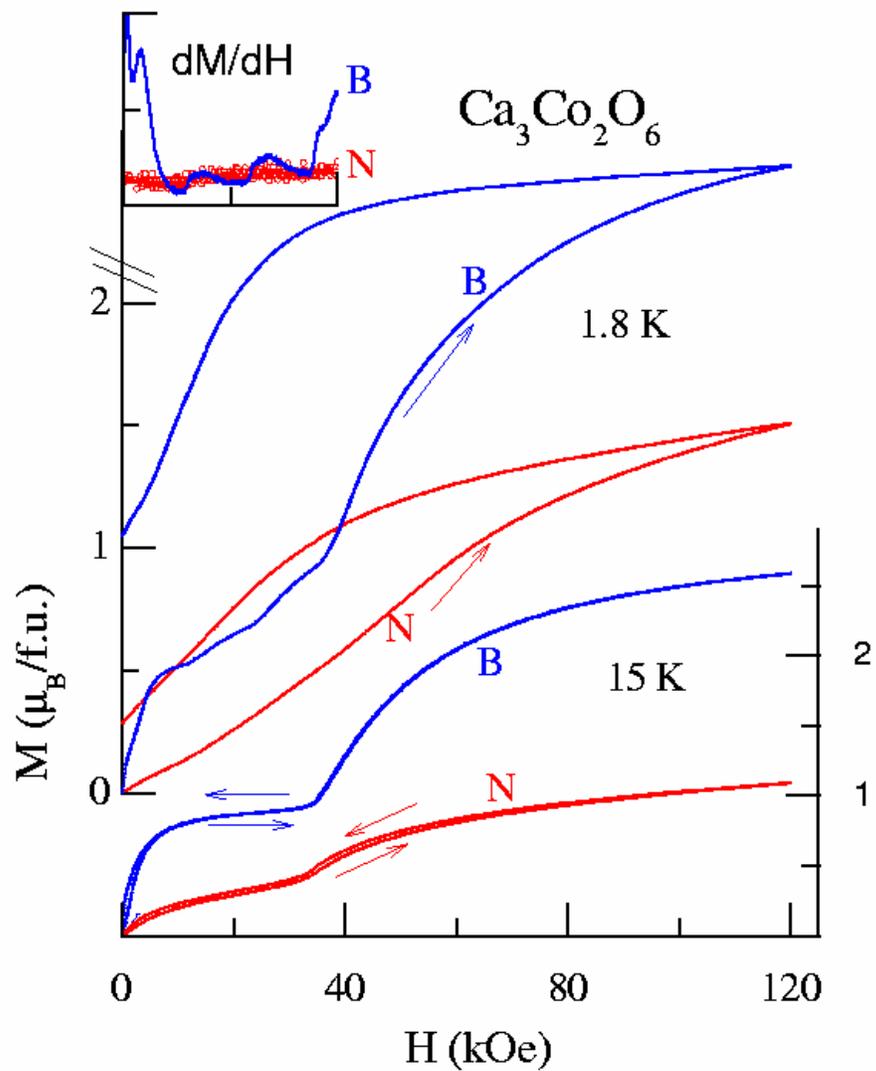

Figure 4:
(color online) Isothermal $M$ at 1.8 and 15 K for the bulk and nanoparticles of $Ca_3Co_2O_6$ for a sweep rate of $H$ of 4 kOe/min. In the inset, the derivative of $M$ below 40 kOe is plotted for both the specimens.

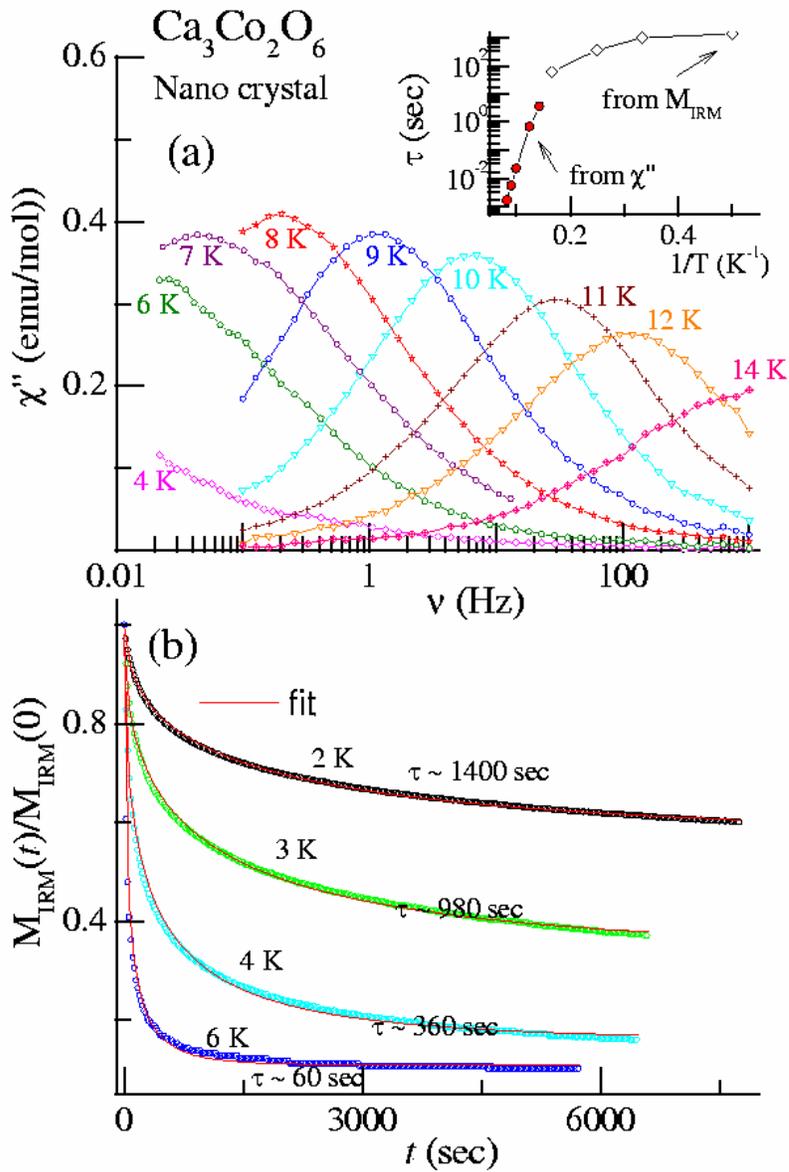

Figure 5:
**(color online) (a)** $\chi''$ as a function of frequency ($\nu$) ($H_{ac}$= 3 Oe) for the nano crystals of $Ca_3Co_2O_6$. The inset shows the relaxation time behavior. The lines through the data points serve as guids to the eyes. **(b)** Normalized isothermal remnant magnetization curves as a function of time.